\documentclass[conference]{IEEEtran}
\IEEEoverridecommandlockouts

\usepackage{amsmath,amsfonts}
\usepackage{algorithmic}
\usepackage{algorithm}
\usepackage{array}
\usepackage[caption=false,font=footnotesize,labelfont=sf,textfont=sf]{subfig}
\usepackage{textcomp}
\usepackage{stfloats}
\usepackage{url}
\usepackage{verbatim}
\usepackage{graphicx}
\usepackage{cite}
\usepackage{color}

\def\BibTeX{{\rm B\kern-.05em{\sc i\kern-.025em b}\kern-.08em
		T\kern-.1667em\lower.7ex\hbox{E}\kern-.125emX}}

\begin{document}

\title{Passive Motion Detection via mmWave Communication System}

\author{Jie Li, Chao Yu, Yan Luo, Yifei Sun and Rui Wang
	\thanks{Jie Li, Chao Yu, Yan Luo, Yifei Sun and Rui Wang are with Department of Electrical and Electronic Engineering, Southern University of Science and Technology (SUSTech), Shenzhen, China (e-mail:$\{$lij2019,yuc2020,luoy2020,sunyf2019$\}$@mail.sustech.edu.cn, wang.r@sustech.edu.cn).
	}
}
\maketitle

\begin{abstract}
	In this paper, an integrated passive sensing and communication system working in 60 GHz band is elaborated, and the sensing performance is investigated in an application of hand gesture recognition. Specifically, in this integrated system, there are two radio frequency (RF) chains at the receiver and one at the transmitter. Each RF chain is connected with one phased array for analog beamforming. To facilitate simultaneous sensing and communication, the transmitter delivers one stream of information-bearing signals via two beam lobes, one is aligned with the main signal propagation path and the other is directed to the sensing target. Signals from the two lobes are received by the two RF chains at the receiver, respectively. By cross ambiguity coherent processing, the time-Doppler spectrograms of hand gestures can be obtained. Relying on the passive sensing system, a dataset of received signals, where three types of hand gestures are sensed, is collected by using Line-of-Sight (LoS) and Non-Line-of-Sight (NLoS) paths as the reference channel respectively. Then a neural network is trained by the dataset for motion detection. It is shown that the classification accuracy rate is high as long as sufficient sensing time is assured. Finally, an empirical model characterizing the relation between the classification accuracy and sensing duration is derived analytically. 
\end{abstract}

\section{Introduction}\label{introduction}

Integrated Sensing and Communication (ISAC) is a promising technology for the sixth generation cellular system, where the sensing capability has potential to assist the wireless communications and provide new services to subscribers \cite{9737357,2107.09574}. Compared with sensing of vehicle's velocity and position, human motion sensing is usually more challenging, as the micro-Doppler effect should be captured for classification. Generally, there are three main approaches of wireless human motion sensing in existing ISAC testbed implementations, including wireless sensing via dedicated waves (e.g.,  Frequency-Modulated Continuous Wave, FMCW), wireless sensing via channel state information (CSI), and passive sensing.

There have been a significant number of works using radar for human motion recognition. For example, in \cite{Adib}, the authors proposed the first coarse multi-person gesture tracking system with FMCW radar, where the direction of a pointing hand can be identified. In \cite{lien2016soli}, Google developed a commercial FMCW-based gesture recognition system, namely the Soli project. In addition, it was shown in \cite{Zhao} that the motion of human skeleton could be reconstructed with recent advances in deep learning. 

Compared with integrated radar and communication systems where dedicated time or frequency resources should be reserved for sensing, there are a number of research works detecting human motion via channel state information (CSI), where the channel sensing does not raise overhead on wireless resource. Particularly, the micro-Doppler effect can also be observed from the variation of the CSI. Moreover, the location of sensing target can be estimated via CSI if multi-antennas and multi-carrier technologies are adopted in the communication system. For example, the CSI was exploited to detect human motions behind the wall in \cite{Seewall}.  In \cite{wang2014eyes}, the authors explored CSI histograms to recognize daily activities, such as cooking in a kitchen and walking from one room to another, based on empirical study. In \cite{wang2015understanding}, the authors proposed theoretical models to derive the relation between CSI and human activities. Recent work \cite{qian2018widar2} jointly estimated the Angle-of-Arrival (AoA), Time-of-Flight (ToF), and Doppler-Frequency-Shift (DFS) for human localization and tracking.

Despite low overhead, CSI-based sensing methods are sensitive to the frequency offset between the transmitter and receiver, which might be a severe problem in millimeter wave (mmWave) band. Although there are methods to eliminate the issue of frequency offset \cite{wang2015understanding}, the performance relies on the propagation path without Doppler shift. On the other hand, passive radar is a sensing approach insensitive to the frequency offset. In \cite{chetty2011through}, passive sensing via WiFi signals was adopted to detect the moving personnel.  It was further shown in \cite{Sun2021TW} that human breathing behind the wall could be detected via passive sensing. Moreover, the CSI-based sensing and passive sensing were compared in \cite{li2020taxonomy}. It was shown that CSI-based system performed better in Line-of-Sight (LoS) scenarios, while passive radar system performed better in Non-Line-of-Sight (NLoS) scenarios.

Although there have been a number of testbeds and experiment results on the passive sensing in sub-6 GHz band, there is still no passive sensing testbed in mmWave band. Note that communications in mmWave band has already been considered in the 5G cellular system and IEEE 802.11ay systems, it is natural to extend the passive sensing technique to mmWave band. Moreover, because of the smaller wavelength, sensing in mmWave band would have higher Doppler resolution.

In this paper, 60 GHz mmWave communication and passive sensing system implementation is elaborated. The 16-antenna phased arrays are deployed at both the transmitter and receiver, so that the mmWave signal transmission can be directed to the sensing target to suppress the interference from LoS path. The performance of the above system is investigated via the exemplary application of hand gesture recognition. Specifically, a dataset of received signals of three gestures is provided and the time-Doppler spectrograms are generated from the dataset. Then Residual Network (ResNet) is trained for classification. Note that the accuracy of classification generally increases with longer sensing duration, as more dynamics of micro-Doppler effects can be captured. Their relation is first tested numerically via ResNet, and then approximated via an analytical expression. It is usually difficult to analyze the performance of ResNet via the statistical learning theory \cite{pml2Book}, the above study provides an empirical model of sensing and learning performance evaluation for ISAC system with a particular classification application.

The remainder of this paper is organized as follows. The model of mmWave passive sensing and communication system is introduced in Section \ref{system-Model}. The signal processing algorithm for passive sensing and the method of classification accuracy approximation are elaborated in Section \ref{signalprocess} and Section \ref{RAM}, respectively. In Section \ref{experiments}, the system implementation is elaborated and the experiment results are demonstrated and discussed. Finally,  the conclusion is drawn in Section \ref{conclusion}.

\section{System Model}\label{system-Model}

\begin{figure}[tb]
	\centering
	\includegraphics[width=\linewidth]{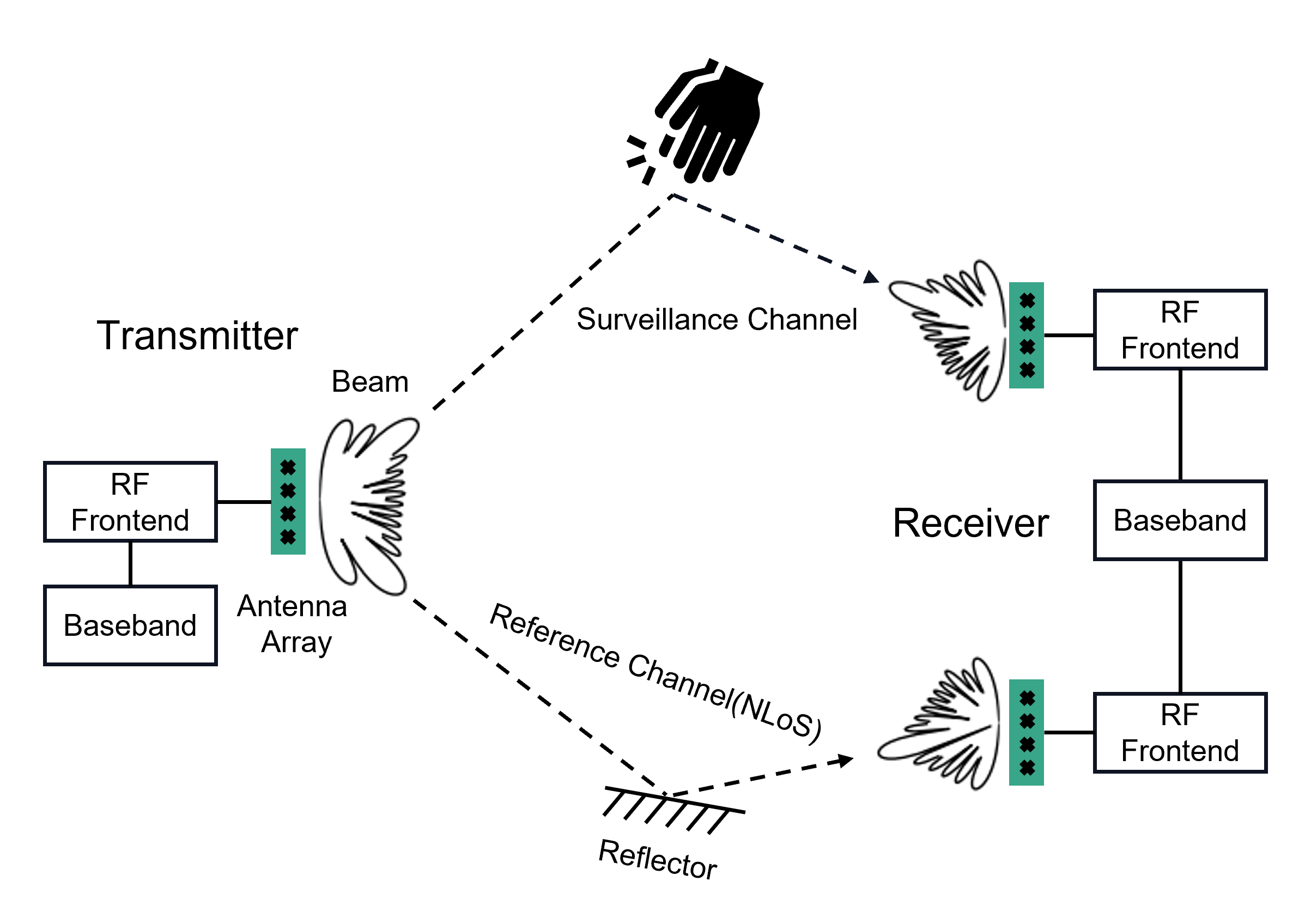}
	\caption{Illustration of the integrated communication and passive sensing system.}
	\label{fig:passiveRadarSystem}
	\vspace{-0.2cm}
\end{figure}

An integrated passive sensing and communication system based on Software-Defined Radio (SDR) and mmWave phased array is illustrated in Fig. \ref{fig:passiveRadarSystem}, which consists of one transmitter and one receiver. Specifically, there is at least one radio frequency (RF) chain at the transmitter and two RF chains at the receiver respectively. Each RF chain is connected with one phased array with $ N $ antenna elements. In order to facilitate simultaneous communication and sensing, both the transmitter and receiver should first estimate the AoDs and AoAs of signal propagation paths and the directions of the sensing target, respectively. In this paper, it is assumed that this angular domain information has already been obtained via the existing approaches \cite{tengwei2017wall}. Hence, the transmitter delivers the information-bearing signals to the receiver by aligning one transmission lobe to static propagation path without Doppler frequency; Meanwhile, it also splits another lobe towards the target human. This can be achieved by separating the $ N $ antenna elements of the phased array into two groups: one group with $ N_d $ elements forms the lobe for data communication, and the remaining $ N_s=N-N_d $ elements form the other lobe for sensing. As illustrated in Fig. \ref{fig:passiveRadarSystem}, the two paths are usually referred to as the reference channel and surveillance channel, respectively. At the receiver, one phased array is used to receive the signal from the reference channel, while the other one is used to collect the echo signal from the surveillance channel.

Specifically, let $ s(t) $ be the information-bearing signal generated at the transmitter, the received signals via the reference and surveillance lobes can be written as 
\begin{equation}\label{eqn:rx-ref}
	y_r (t) = \sum_{i=1}^{L_r} \alpha_{r}^i s(t-\tau_r^i)  e^{-j2\pi \Delta t} + n_r(t)
\end{equation}
and
\begin{equation}\label{eqn:rx-sur}
	y_s (t) = \sum_{i=1}^{L_{s}} \alpha_{s}^i  s(t-\tau_s^i)e^{-j2\pi f_i t} e^{-j2\pi \Delta t}+ n_s(t),
\end{equation}
respectively. In (\ref{eqn:rx-ref}), $ L_r $ denotes the number of resolvable paths received by the RF chain for the reference channel, $ \alpha_{r}^i $ and $ \tau_r^i $ denote the gain and delay of the $ i $-th path respectively, $ \Delta $ is the frequency offset between the transmitter and the receiver, and $ n_r(t) $ is the noise. Similarly in (\ref{eqn:rx-sur}), $ L_s $ denotes the number of resolvable paths received by the RF chain for the surveillance channel, $ \alpha_{s}^i $, $ \tau_s^i $ and $f_i$ denote the gain, delay and Doppler shift of the $ i $-th path respectively. As a remark notice that the LoS path or NLoS path via static reflectors is selected as the reference channel, so that there is no Doppler shift in (\ref{eqn:rx-ref}). Moreover, with highly directional beams at both the transmitter and the receiver, there is usually one dominant path in $ y_r(t) $. Hence, (\ref{eqn:rx-ref}) can be rewritten as
\begin{equation}
	y_r (t) = \alpha_{r}^1 s(t-\tau_r^1)  e^{-j2\pi \Delta t} + \widetilde{n}_r(t),
\end{equation}
where received signals from other paths are merged into the noise $ \widetilde{n}_r(t) $.

\vspace{-0.2cm}
\section{Signal Processing of Passive Sensing}\label{signalprocess}

Given the received signals from reference channel and surveillance channel, the Doppler shifts raised by human motion, denoted as $\{f_i|i=1,\ldots,L_s\}$, can be estimated via the following cross ambiguity function,
\begin{equation}\label{eqn:caf}
	R(\tau,f) = \int_{0}^{T_w} y_s(t) y_r^{\dagger}(t-\tau) e^{j 2 \pi f  t} dt, 
\end{equation}
where $(.)^{\dagger}$ denotes the complex conjugate, and $T_w$ is the coherent integration time (CIT). It can be observed that a local peak value of $ R(\tau,f) $ can be achieved when $ \tau $ and $ f $ match the delay differences and Doppler shifts between $ y_d(t) $ and $ y_s(t) $. 

Note that in the scenario of human motion sensing, the number of paths, path delays and Doppler shifts, i.e., $L_s$, $ \tau_s^i $ and $f_i$ in (\ref{eqn:rx-sur}), are all time varying. The calculation of cross ambiguity function over the CIT as (\ref{eqn:caf}) will mix the Doppler shits of different time instances. Similar to the works on sub-6 GHz band \cite{li2020passive,li2020taxonomy}, a sliding window with length of CIT is applied on the above cross ambiguity function to generate the time-Doppler spectrogram,
\begin{equation}\label{eqn:cafs}
	\widetilde{R}(f,t) = \max_{\tau} \int_{t}^{t+T_w} y_s(x) y_r^{\dagger}(x-\tau) e^{j 2 \pi f  x} dx,
\end{equation}
where $T_w$ is the length of sliding window. Note that larger $ T $ will lead to better Doppler resolution, but mixture of time-varying Doppler shifts and higher computation complexity. Since we focus on the feature extraction of Doppler shifts in this work, the delay $\tau$ is maximized in \eqref{eqn:cafs}.

\begin{figure}[tb]
	\centering
	\includegraphics[width=0.8\linewidth]{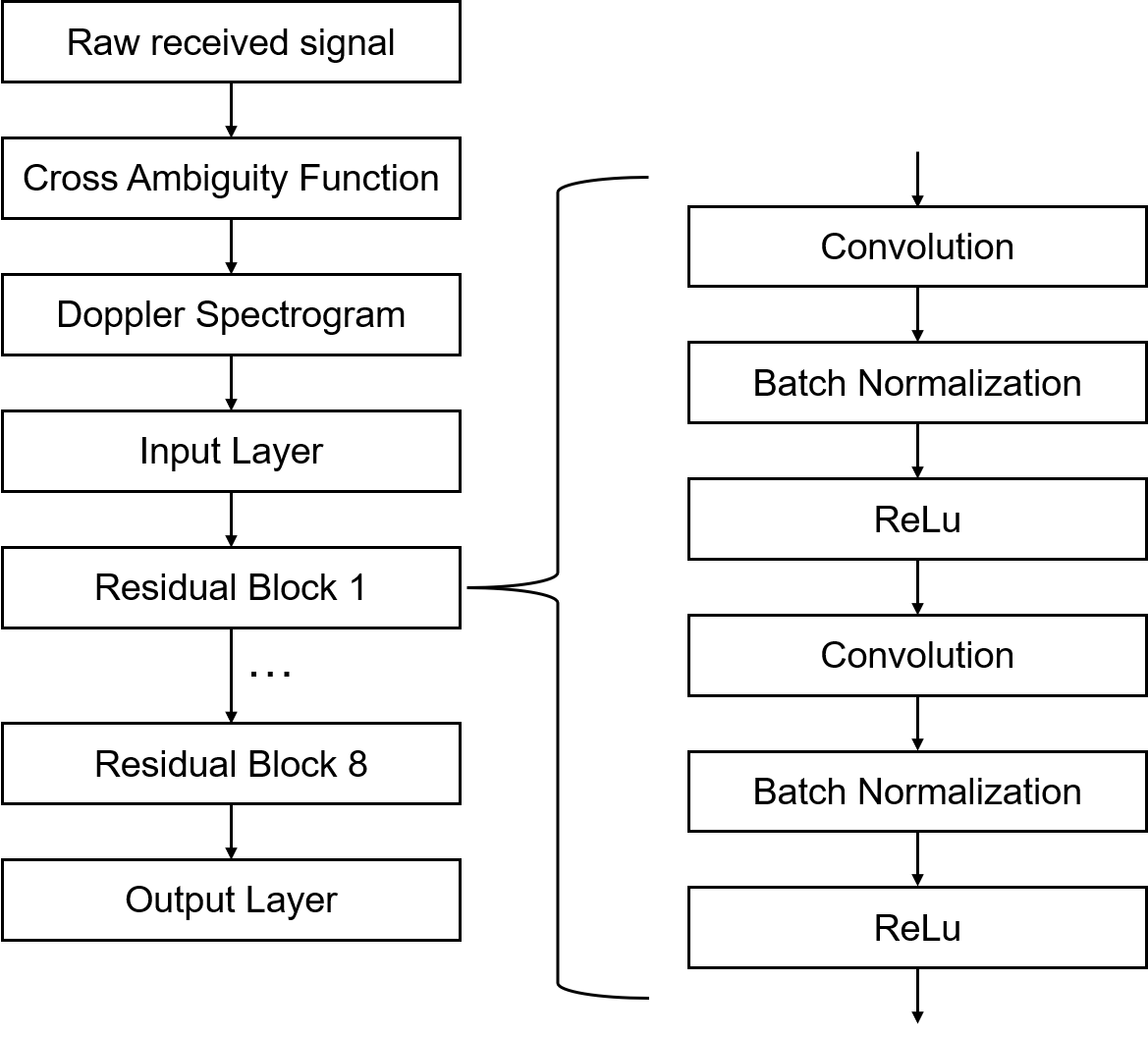}
	\caption{Training procedure for motion detection.}
	\label{fig:flow}
	\vspace{-0.4cm}
\end{figure}

In this paper, the sensing signals for different hand gestures are collected. To classify the gestures according to the time-Doppler spectrogram, a type of ResNet, namely ResNet-18 \cite{He2016resnet}, is adopted, as illustrated in Fig. \ref{fig:flow}. The input of the ResNet is the time-Doppler spectrogram obtained in (\ref{eqn:cafs}), and its output is the classified gesture category. The ResNet includes eight residual blocks and one fully connected layer. Each residual block contains two convolutional layers, two batch normalization layers, and two ReLU layers.

\section{Classification Accuracy Model}\label{RAM}
It is of significant interests to investigate the relation between motion classification accuracy, denoted as $\psi$, and sensing duration $T$ in ISAC scheduler design. Note that it is usually difficult to investigate the performance of ResNet analytically, the approximation method proposed in \cite{zhou2020learning,domhan2015speeding} is adopt in this paper. Let $\psi_i\ (i=1,2,...,Q)$ be the tested classification accuracy with sensing durations $T_i\ (i=1,2,...,Q)$, respectively. The relation between $\psi$ and $T$ can be approximated via the following expression,

\begin{equation}\label{eqn:curve}
	\Psi = \gamma - \alpha T ^{-\beta},
\end{equation}
where $\gamma$, $\alpha$, and $\beta$ are the tuning paramters. The values of $\gamma$, $\alpha$ and $\beta$ can be obtained via the following optimization problem.
\begin{equation}\label{eqn:optimization}
	\mathop{\arg\max}_{\gamma,\alpha,\beta} \frac{1}{Q} \sum_{i=1}^{Q} \psi _i - (\gamma - \alpha T_i ^{-\beta} ).
\end{equation}

\section{Experiments}\label{experiments}
\subsection{Implementation}

\begin{figure}[tb]
	\centering
	\includegraphics[width=\linewidth]{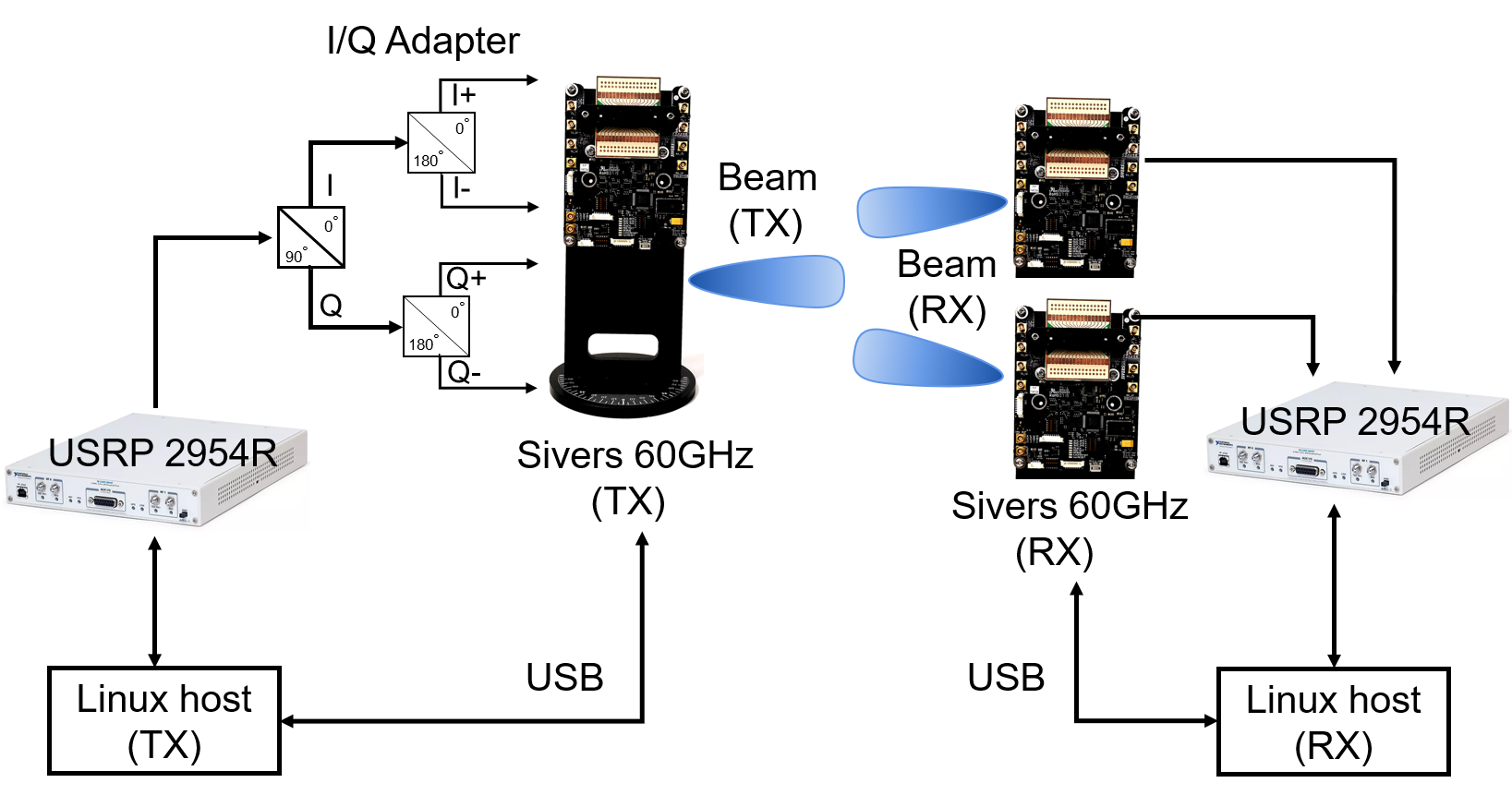}
	\caption{Block diagram of system implementation.}
	\label{fig:blockDiagram}
	\vspace{-0.4cm}
\end{figure}

The overall system diagram is illustrated in Fig. \ref{fig:blockDiagram}. At the transmitter, one NI USRP-2954R \cite{USRP-X310} is adopted to generate an intermediate frequency (IF) signal centered at $ 500 $ MHz, which is further upconverted to 60 GHz and transmitted by one Sivers 60 GHz phased array \cite{evk}. At the receiver, two phased arrays are connected with one USRP to receive the signals from the reference channel and surveillance channel respectively. All the phased arrays are controlled by laptops, so they can switch beams collaboratively.  

The transmission signal $ s(t) $ consists of a training sequence with a duration of $ 16 $ us, followed by OFDM-modulated data payload with a duration of $ 200 $ us. From the training sequence, the CSI of surveillance channel can also be obtained. Hence the time-Doppler spectrogram generated from CSI is also obtained for comparison. As a remark notice, it is not necessary for the receiver to estimate the CSI, if only passive sensing is considered.

\begin{figure}[tb]
	\centering
	\includegraphics[width=0.8\linewidth]{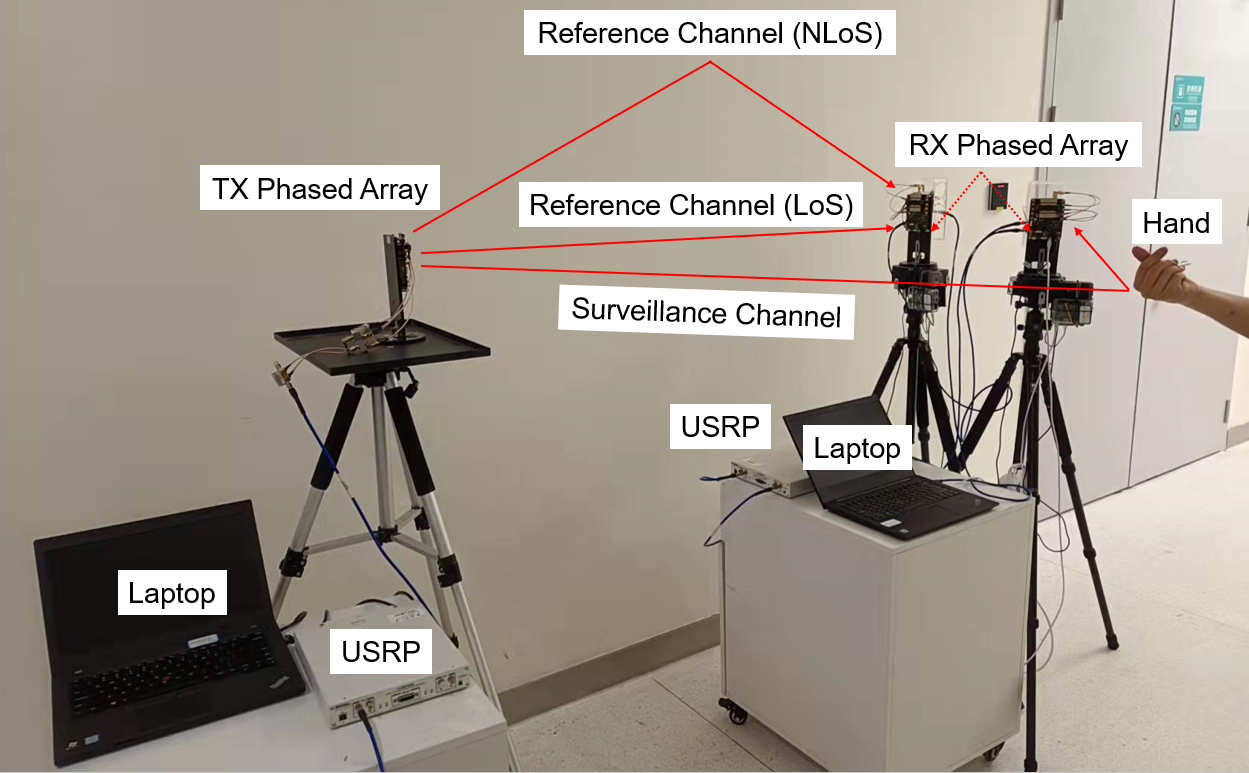}
	\caption{Experiment Layout.}
	\label{fig:layout}
	\vspace{-0.4cm}
\end{figure}

As illustrated in Fig. \ref{fig:layout}, two scenarios of passive sensing are considered in the experiment, namely LoS and NLoS scenarios. In the LoS scenario, the LoS path between the transmitter and receiver is considered as the reference channel; Whereas, the NLoS path via wall reflection is used as the reference channel in the NLoS scenario. Since the LoS may be blocked frequently in practice, sensing robustness can be enhanced by switching to NLoS paths as the reference channel. 

There are three types of gestures to be sensed in the experiment, including pushing hand, thumb adduction, and rubbing fingers. Each hand gesture is sampled via the passive sensing system for $ 100 $ times in both LoS and NLoS scenarios, so that a dataset is obtained. In each sample, the information-bearing signal $ s(t) $ is transmitted for $ 9000 $ times. For ambiguity processing, the window duration is $ T_w = 0.1 $ s.

\subsection{Time-Doppler Spectrograms}

\begin{figure}[tb] 
	\centering
	
	\subfloat[]{
		\label{fig:spectrogramA} 
		\includegraphics[width=0.45\linewidth]{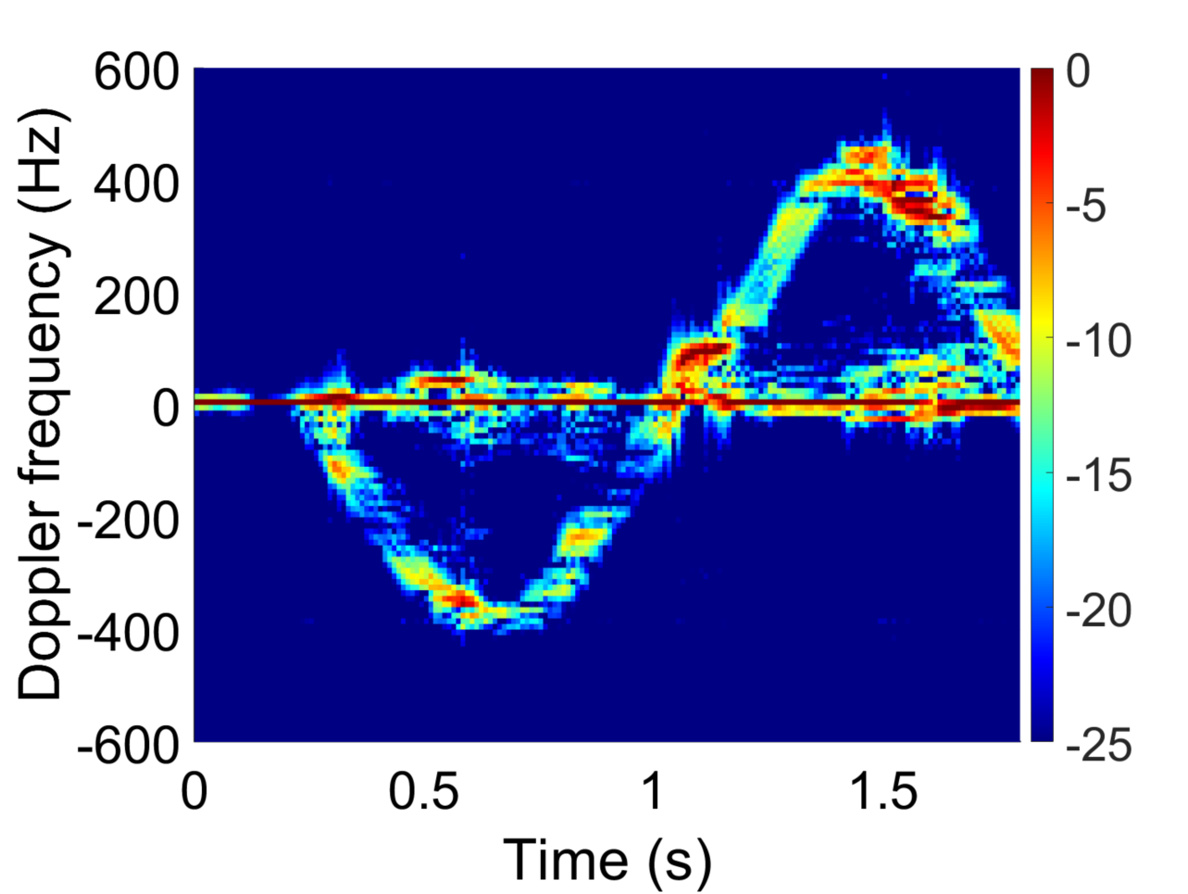}}
	\subfloat[]{
		\label{fig:spectrogramB}
		\includegraphics[width=0.45\linewidth]{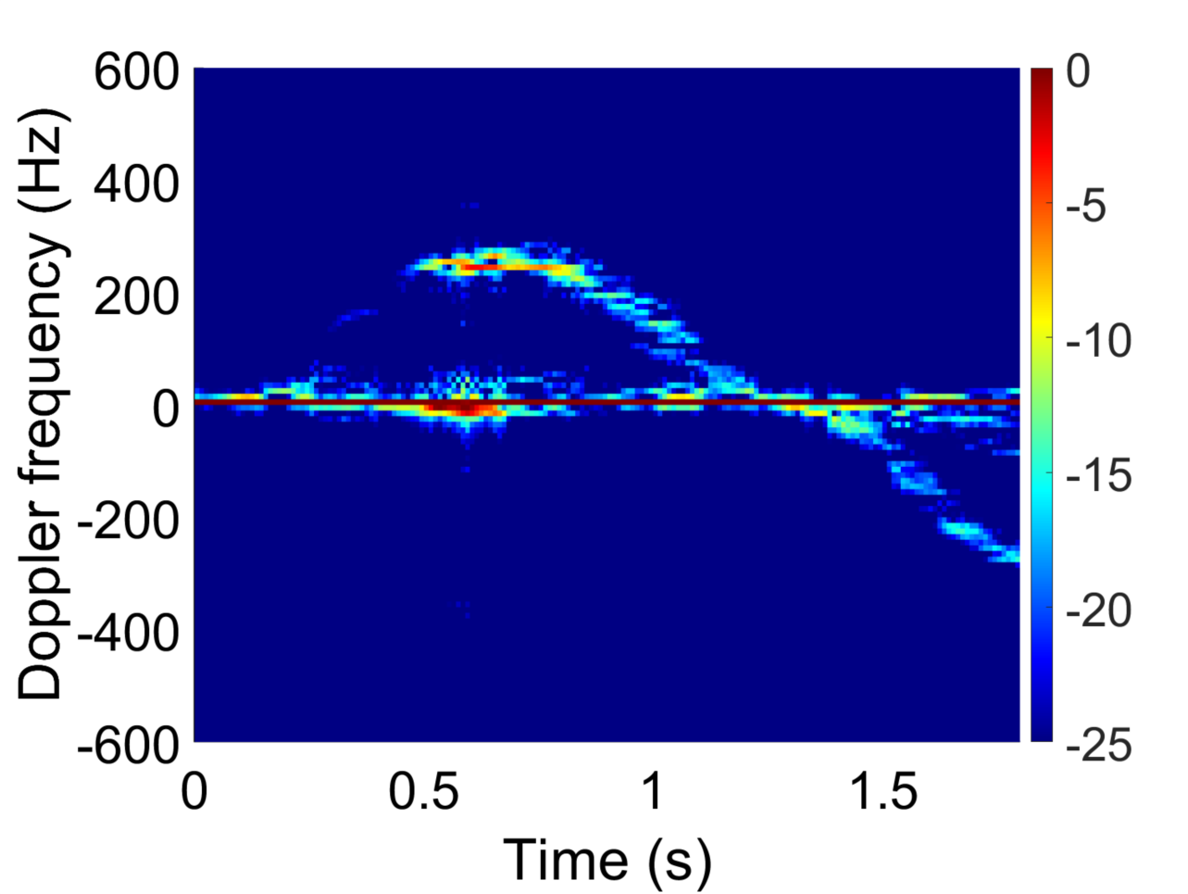}}
	\vspace{-0.4cm}
	\newline
	\subfloat[]{
		\label{fig:spectrogramC} 
		\includegraphics[width=0.45\linewidth]{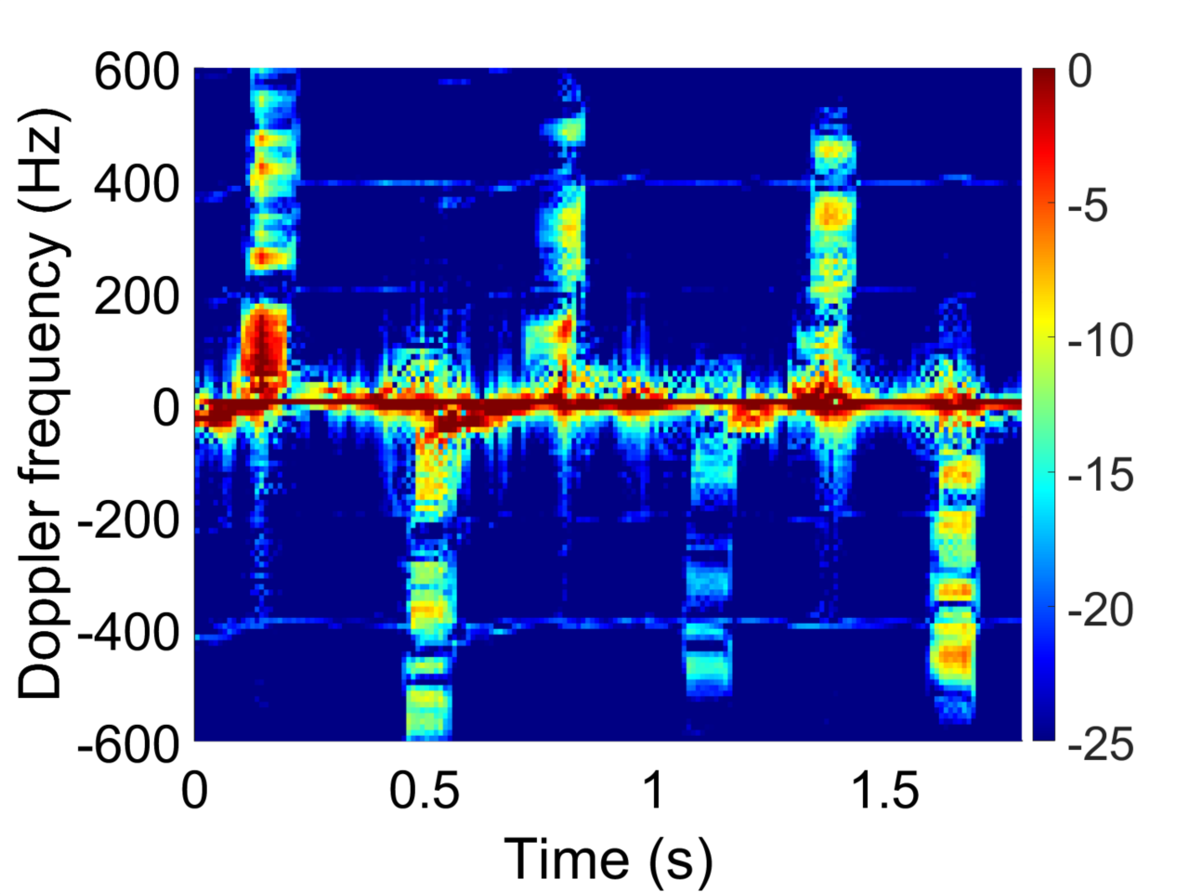}}
	\subfloat[]{
		\label{fig:spectrogramD}
		\includegraphics[width=0.45\linewidth]{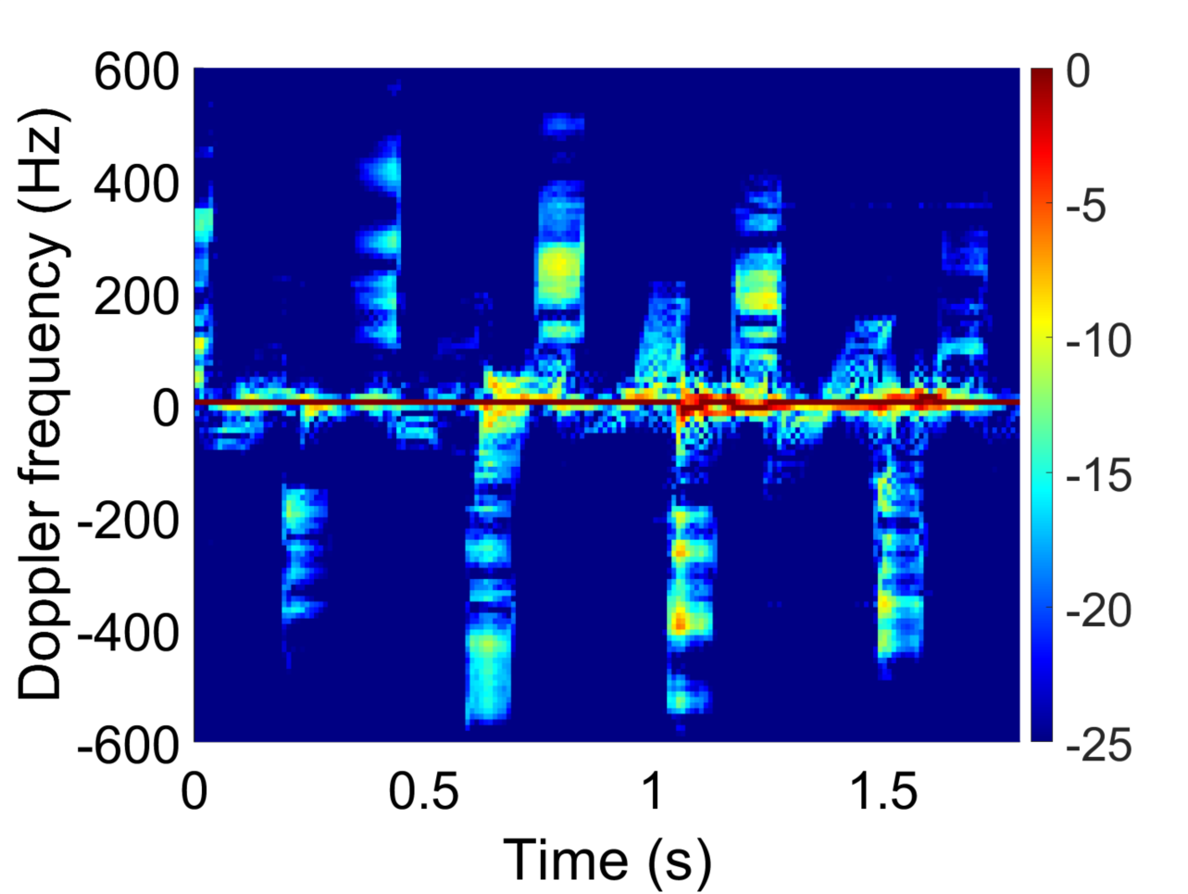}}
	\vspace{-0.4cm}
	\newline
	\subfloat[]{
		\label{fig:spectrogramE} 
		\includegraphics[width=0.45\linewidth]{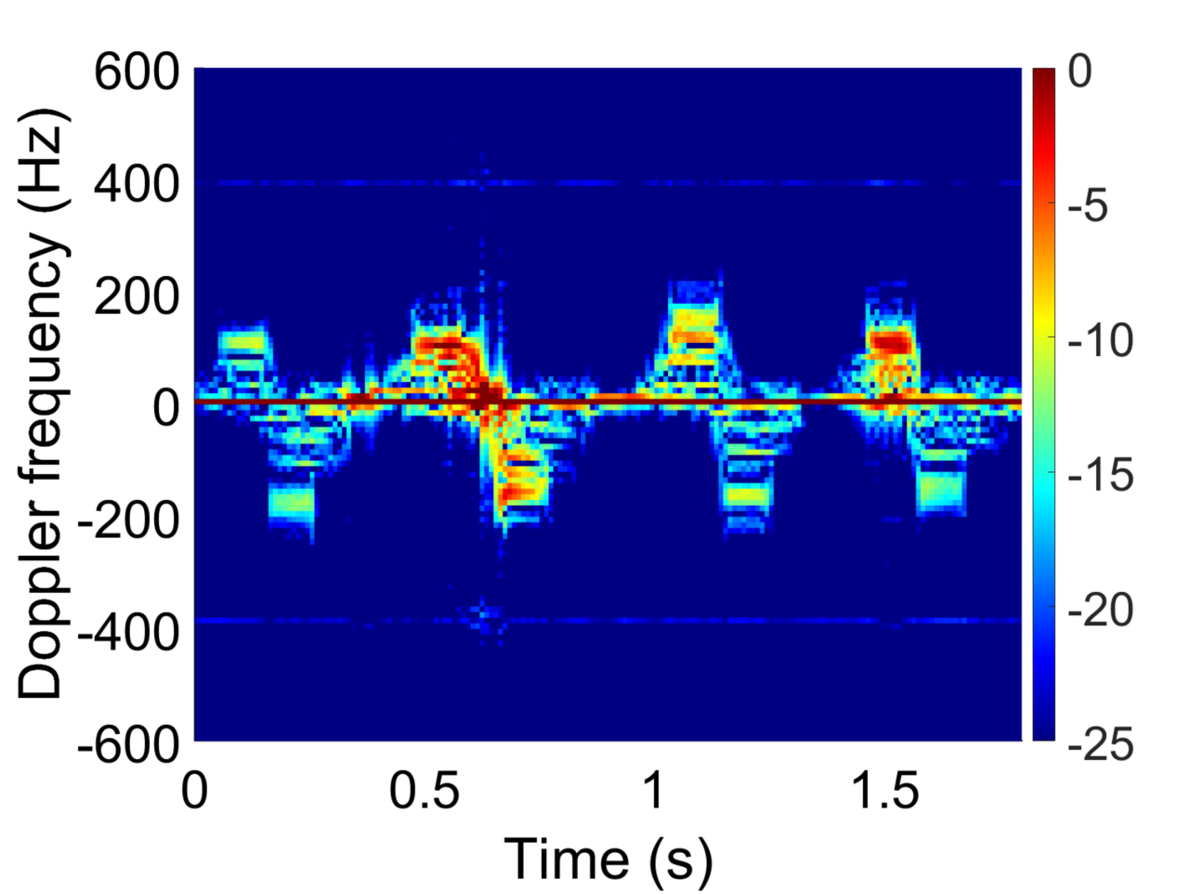}}
	\subfloat[]{
		\label{fig:spectrogramF}
		\includegraphics[width=0.45\linewidth]{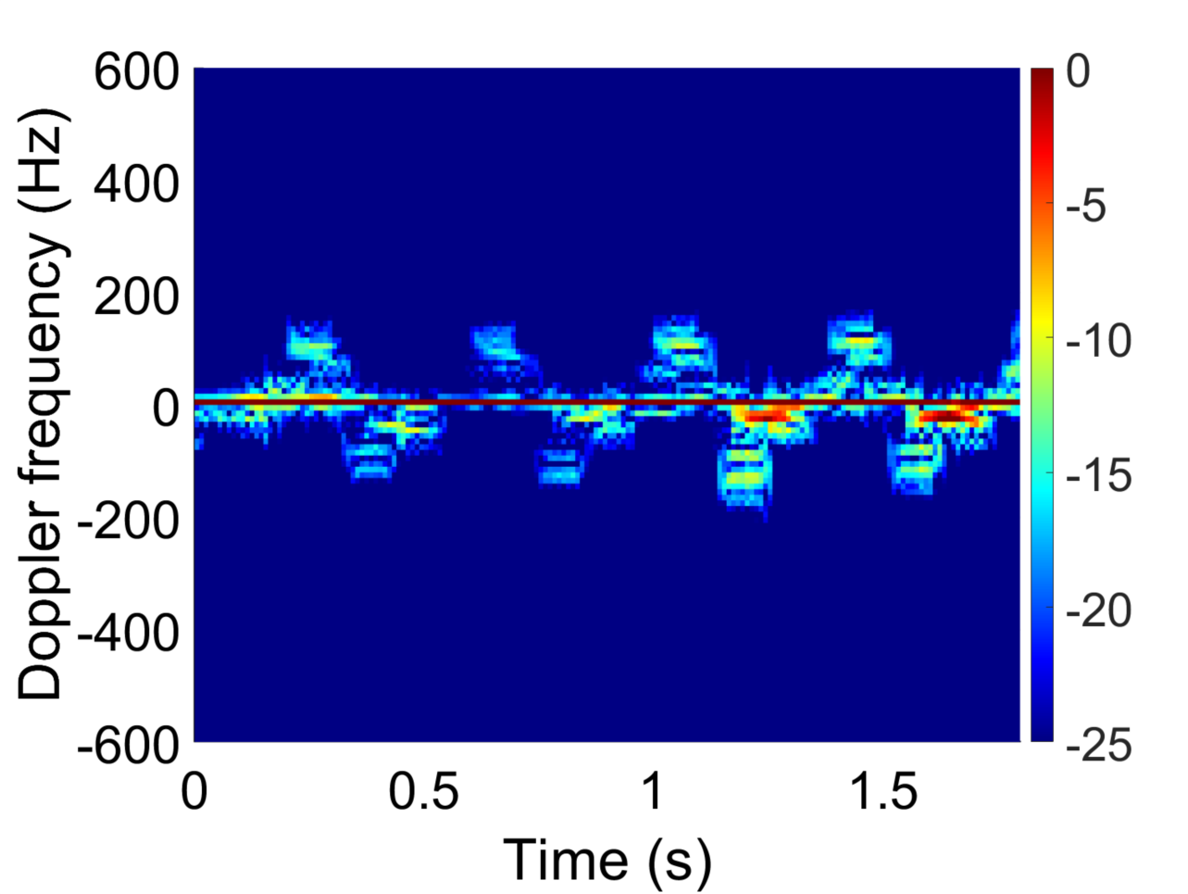}}
	\vspace{-0.1cm}
	
	\caption{Spectrograms obtained from  LoS scenario: (a) pushing hand, (c) thumb adduction, (e) rubbing finger, and from NLoS scenario:  (b) pushing hand, (d) thumb adduction, (f) rubbing finger. }
	\label{fig:spectrogram}
	\vspace{-0.4cm}
\end{figure}

The examples of time-Doppler spectrogram for the three gestures in LoS scenario are illustrated in Fig. \ref{fig:spectrogram}(a)(c)(e). It can be observed that the three gestures have significantly different patterns of time-Doppler spectrograms. For example, pushing hand leads to smooth variation of Doppler frequency between positive and negative peaks, while thumb adduction leads to sharp impulses of Doppler frequency. The peak Doppler frequencies of all three gestures are different. For example, the Doppler frequency generated by rubbing finger is less significant compared with the other two gestures. This is because of the lower amplitude of finger motion. Hence, it is feasible to differentiate the three gestures via their time-Doppler spectrograms. Moreover, the zero Doppler frequency component in the spectrogram demonstrates the existence of static scatters in the surveillance channel. 

The examples of time-Doppler spectrogram in NLoS scenario are illustrated in Fig. \ref{fig:spectrogram}(b)(d)(f). Because the NLoS signal is weaker than that in LoS scenario, the Doppler frequency illustrated in NLoS scenario is weaker than that in LoS scenario in general. However, it is still sufficiently clear to distinguish the three different gestures.  This demonstrates the feasibility of using NLoS path as the reference channel. Note that the NLoS reference channel is helpful especially when the LoS path is blocked or the sensing target is close to the LoS path. 

\begin{figure}[tb] 
	\centering
	\subfloat[]{
		\label{fig:csispectrogramA} 
		\includegraphics[width=0.45\linewidth]{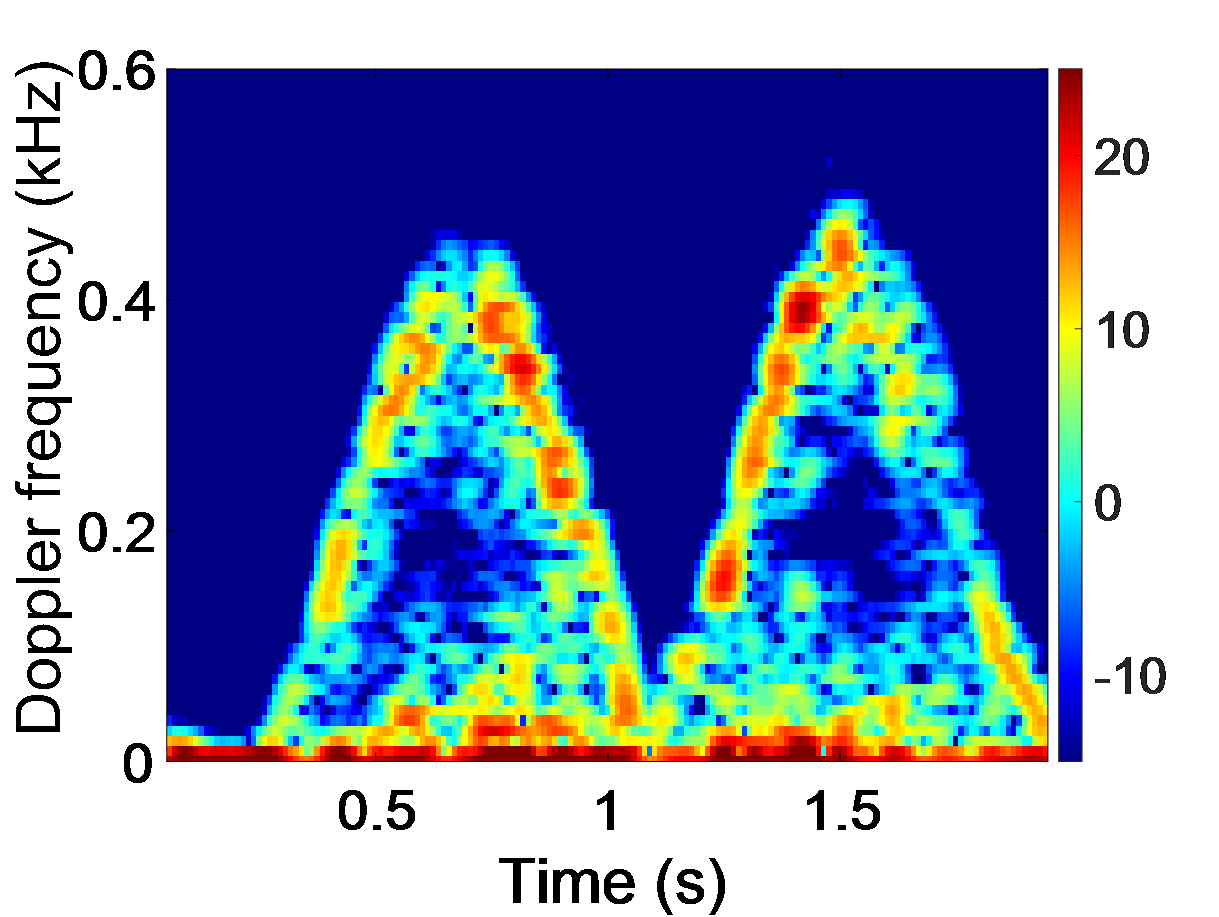}}
	\hfill
	\subfloat[]{
		\label{fig:csispectrogramB}
		\includegraphics[width=0.45\linewidth]{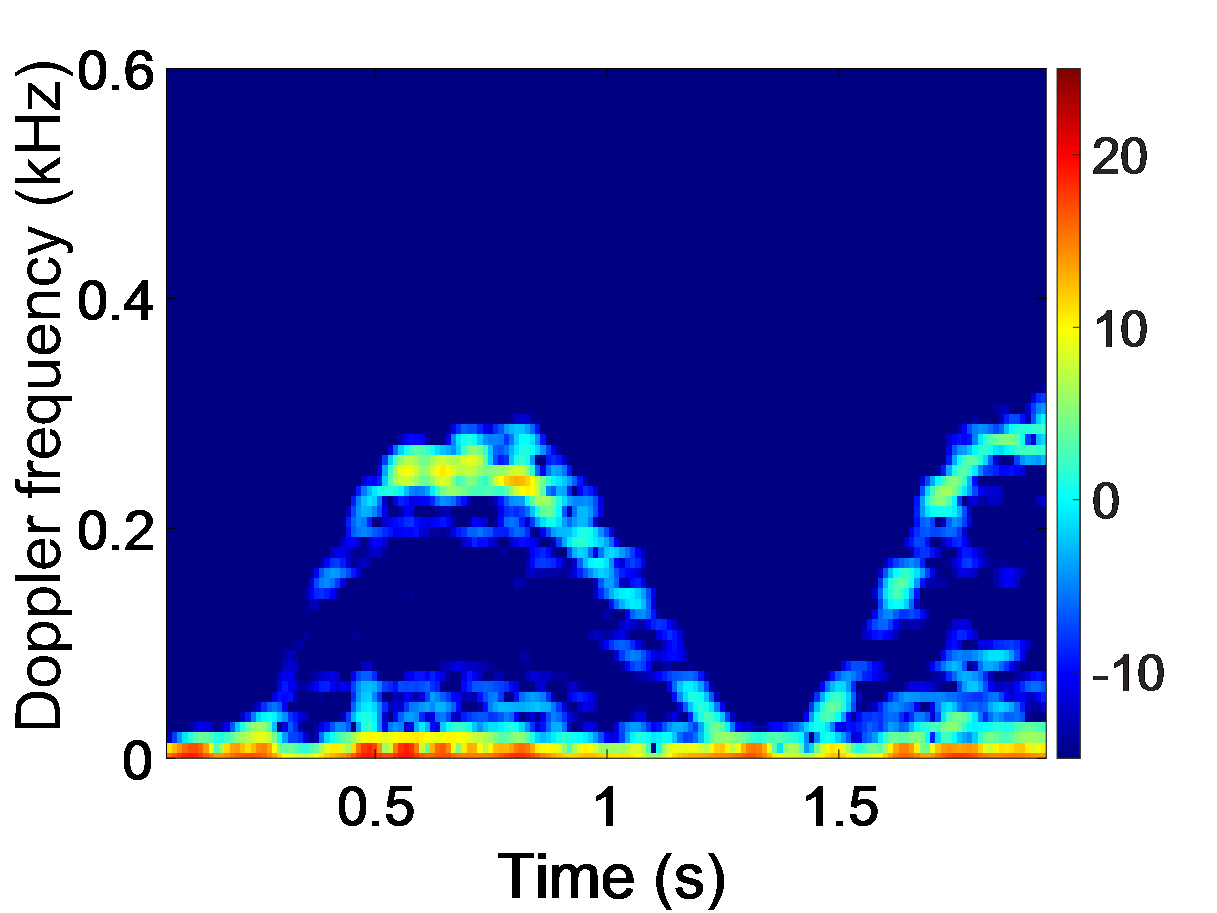}}
	\vspace{-0.4cm}
	\hfill
	\newline
	\subfloat[]{
		\label{fig:csispectrogramC} 
		\includegraphics[width=0.45\linewidth]{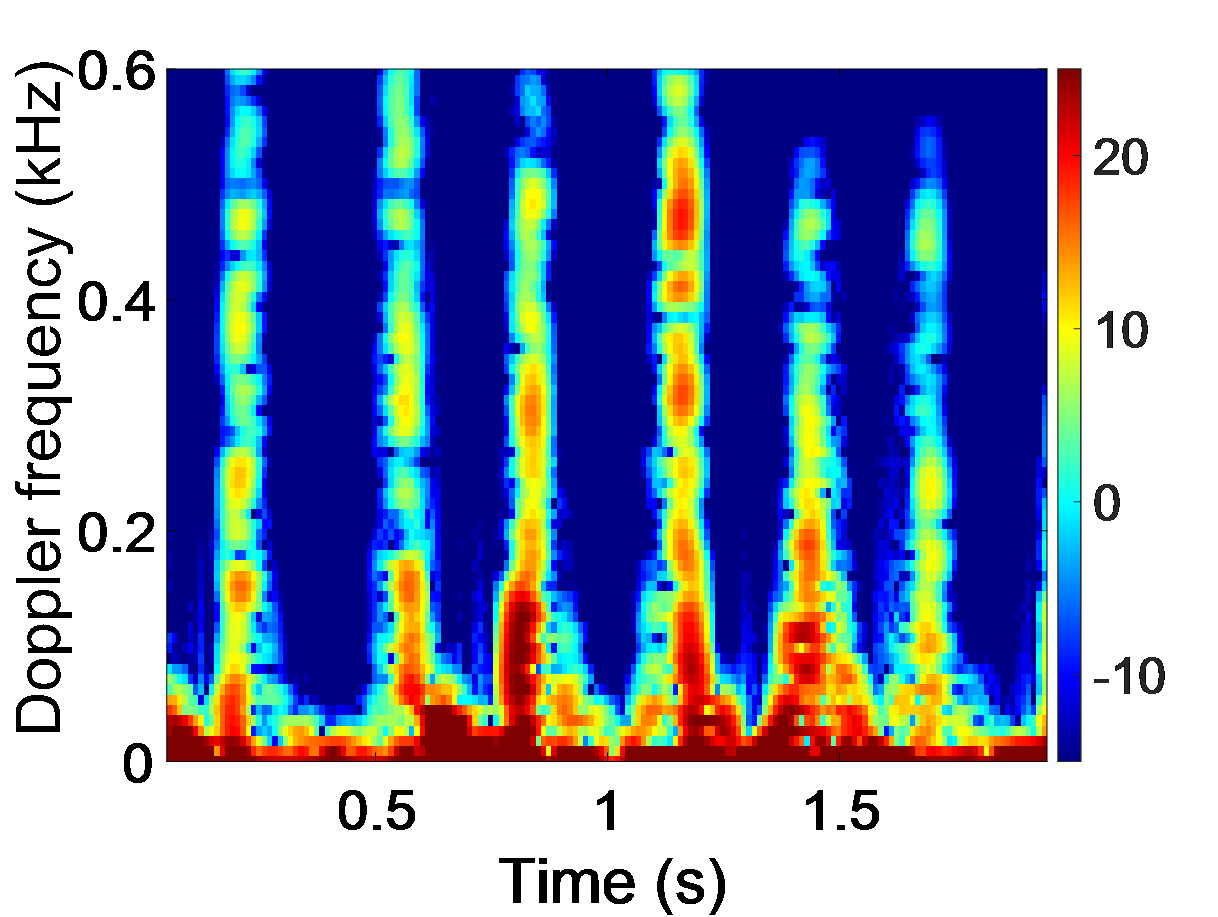}}
	\hfill
	\subfloat[]{
		\label{fig:csispectrogramD}
		\includegraphics[width=0.45\linewidth]{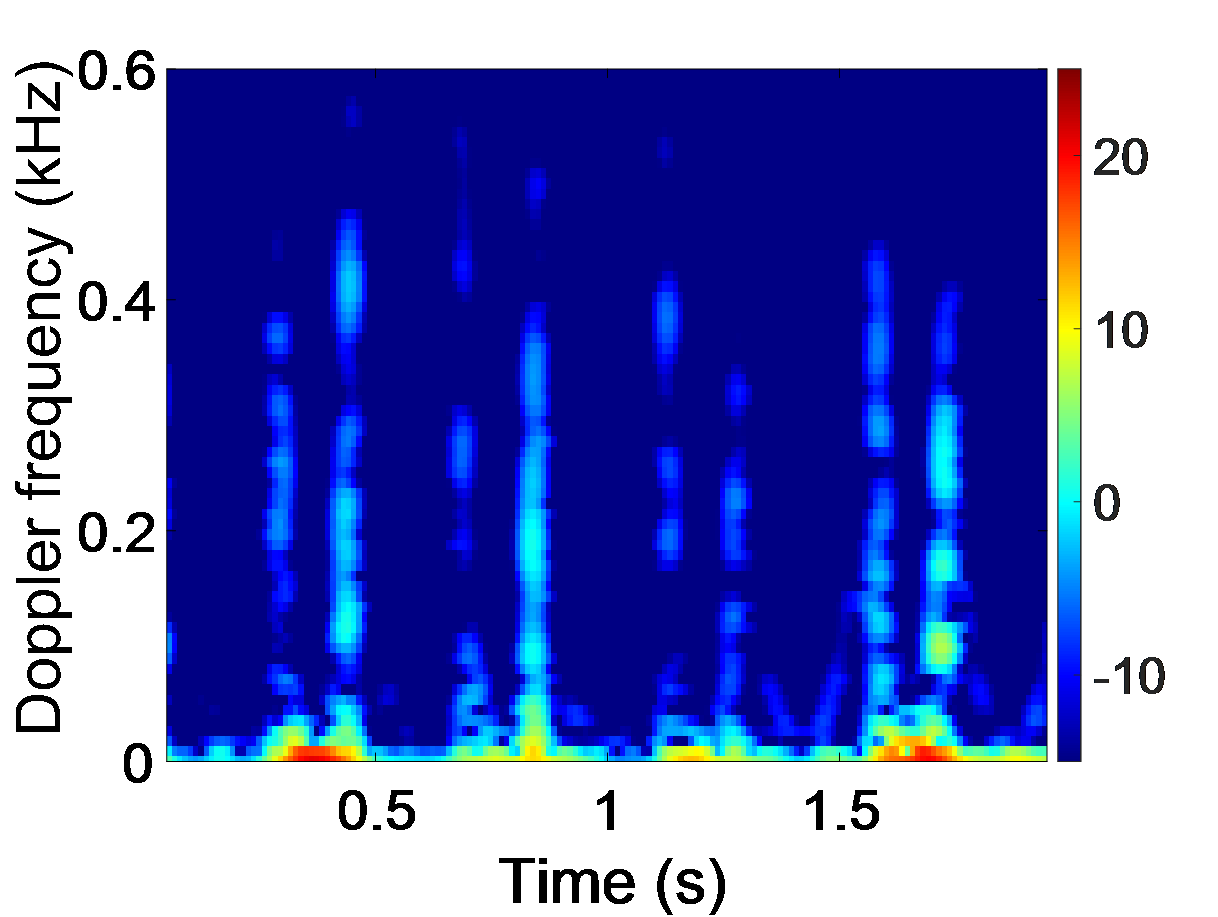}}
	\vspace{-0.4cm}
	\hfill
	\newline
	\subfloat[]{
		\label{fig:csispectrogramE} 
		\includegraphics[width=0.45\linewidth]{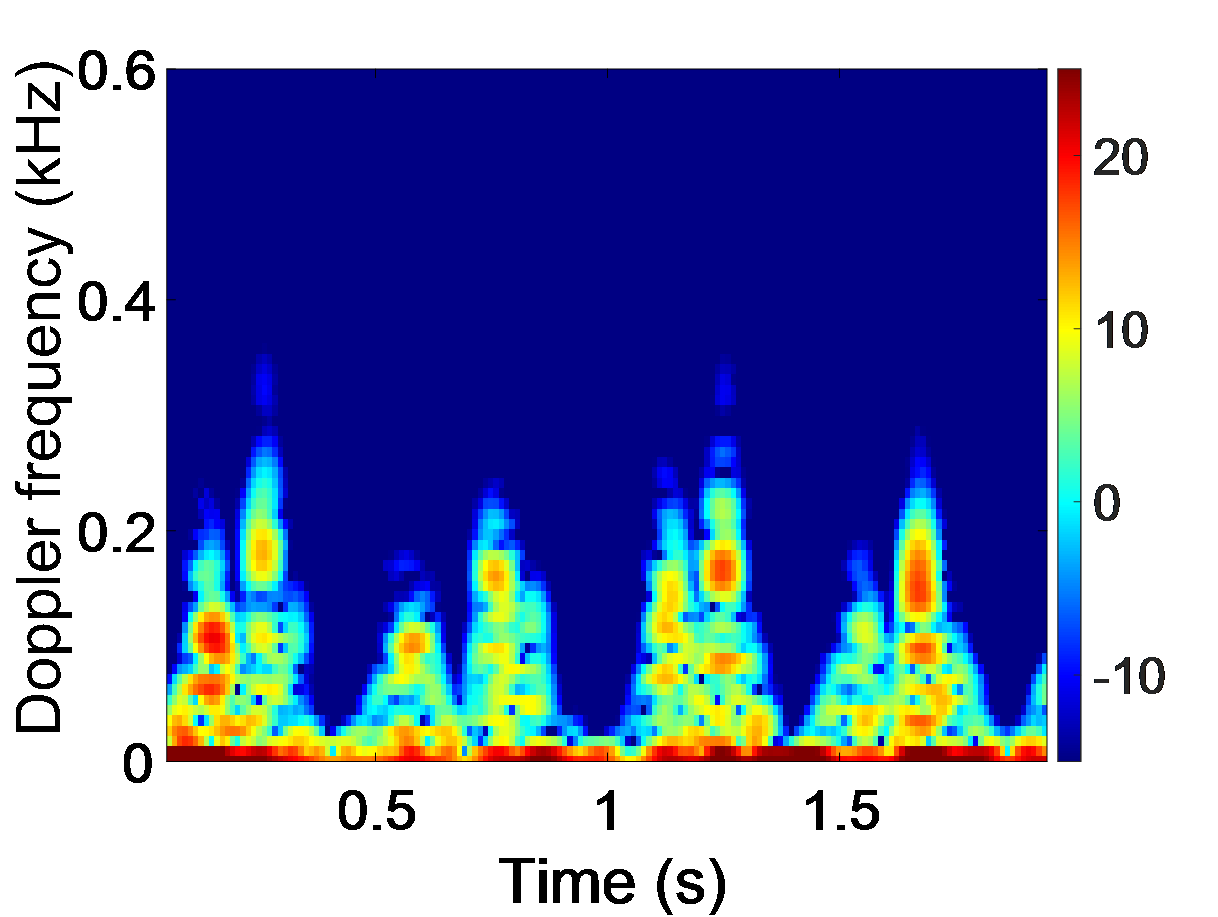}}
	\hfill
	\subfloat[]{
		\label{fig:csispectrogramF}
		\includegraphics[width=0.45\linewidth]{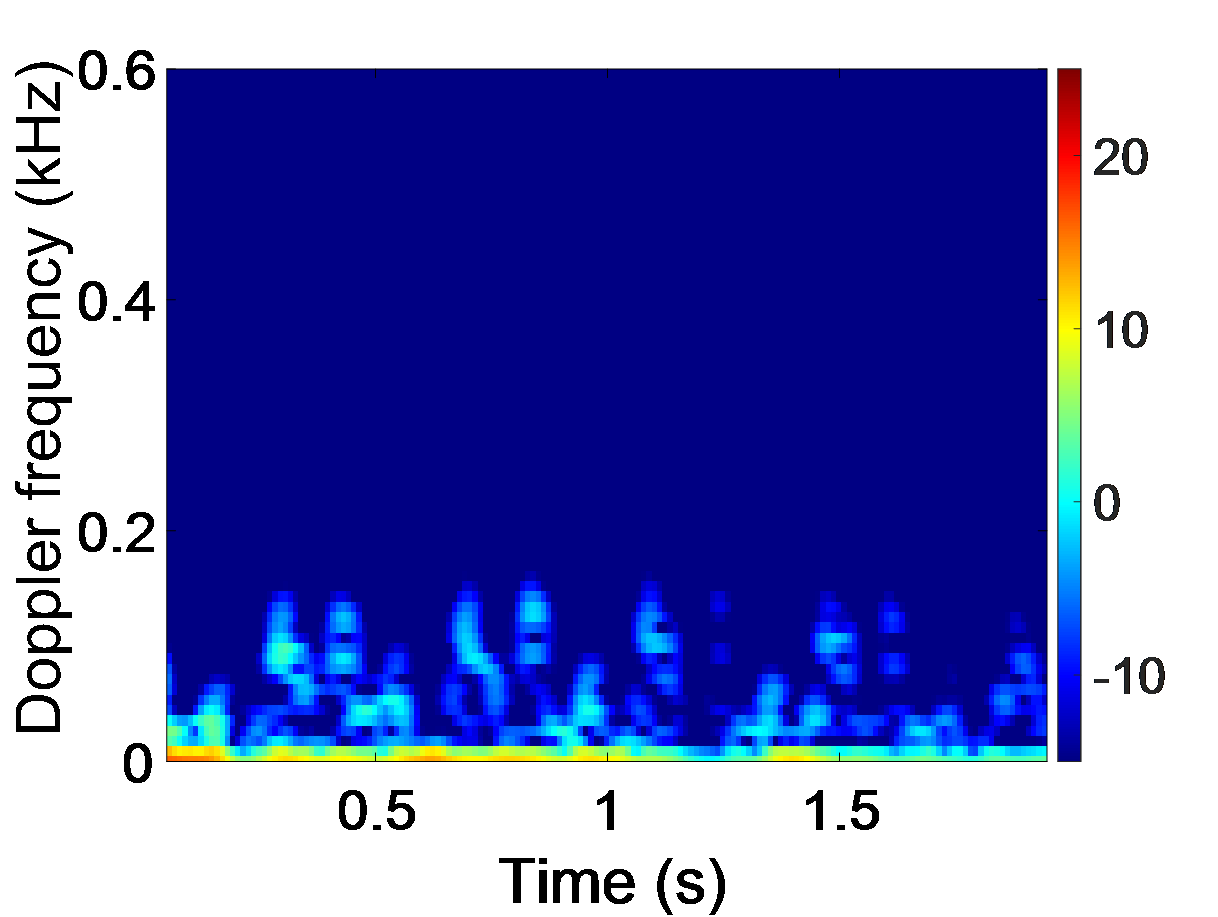}}
	\vspace{-0.1cm}
	
	\caption{CSI spectrograms obtained from  LoS scenario: (a) pushing hand, (c) thumb adduction, (e) rubbing finger, and from NLoS scenario:  (b) pushing hand, (d) thumb adduction, (f) rubbing finger. }
	\label{fig:csispectrogram} 
	\vspace{-0.4cm}
\end{figure}

In Fig. \ref{fig:csispectrogram}, the time-Doppler spectrograms of CSI of the surveillance channel are illustrated for comparison. As a remark notice that the estimation of surveillance channel is necessary to generate Fig. \ref{fig:csispectrogram}. However, the frame synchronization and channel estimation are not required in passive sensing. Since the transmitter and the receiver are not well synchronized, we only show the spectrogram of CSI's magnitude, where the carrier frequency offset between the transmitter and receiver can be eliminated as explained in \cite{wang2015understanding}. Compared with the spectrograms in Fig. \ref{fig:spectrogram}, the spectrogram of CSI's magnitude is different in the following aspects: (1) It only has non-negative Doppler frequency components; (2) It consists of not only the actual Doppler frequencies of the prorogation paths in surveillance channel, but also their mutual couplings \cite{wang2015understanding}. The latter effect can be observed by comparing the spectrograms in Fig. \ref{fig:spectrogram}(a) and Fig. \ref{fig:csispectrogram}(a), where Doppler frequencies due to mutual coupling can be found in Fig. \ref{fig:csispectrogram}(a).  Moreover, the spectrogram of  CSI's magnitude relies on the existence of static propagation path, as elaborated in \cite{li2020taxonomy}. When the  static propagation path is weak, the spectrogram may not be able to show the correct Doppler frequency, which can be observed in Fig. \ref{fig:csispectrogram}(d) and Fig. \ref{fig:csispectrogram}(f). 

\subsection{Motion Detection}
\begin{figure}[tb]
	\centering
	\includegraphics[width=\linewidth]{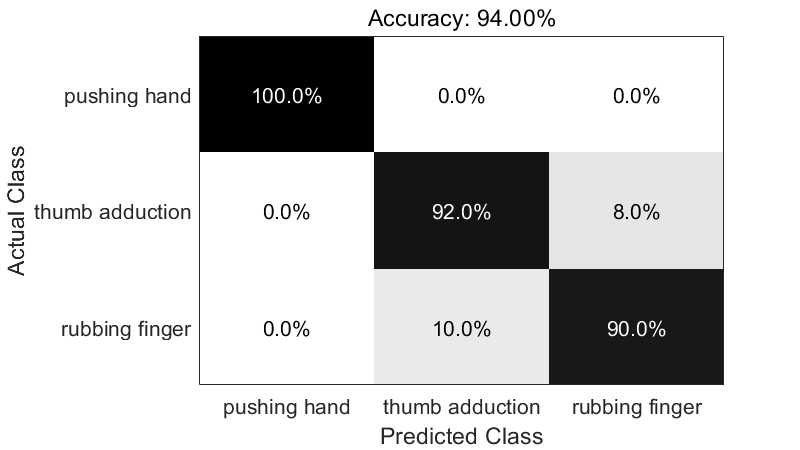}
	\caption{Classification matrix.}
	\label{fig:ConfMatrix}
	\vspace{-0.4cm}
\end{figure}

In order to classify the three gestures via the ResNet, we train the network with 150 samples (50 samples per gesture) and the mini-batch size of 16. The remaining samples are used as the test set. The classification results are shown in Fig. \ref{fig:ConfMatrix}. With a sensing duration $T\!=\!2 $ s, the overall  accuracy of NLoS scenario is $ 94 $\%, where the classification accuracy of pushing hand is 100\%. This is because the spectrogram of pushing hand is distinct from those of thumb adduction and rubbing finger. Moreover, the classification accuracies of thumb adduction and rubbing finger are also above 90\%, which demonstrates good performance of gesture recognition.

\begin{figure}[tb]
	\centering
	\includegraphics[width=0.8\linewidth]{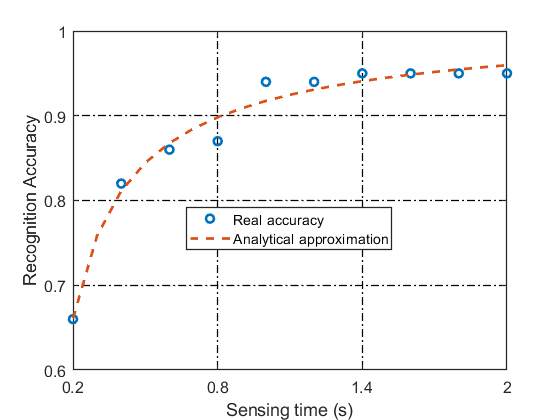}
	\caption{Classification accuracy versus the sensing time.}
	\label{fig:recgAccuracy}
	\vspace{-0.4cm}
\end{figure}

The classification accuracy versus the sensing duration is illustrated in Fig. \ref{fig:recgAccuracy}. It can be observed that the classification accuracy increases with respect to the length of sensing time. This is because longer sensing time could capture more dynamics in micro-Doppler effect, and thus better performance.

According to Section \ref{RAM}, it can be calculated that the optimal values of $\gamma$, $\alpha$ and $\beta$ for classification accuracy approximation are $1.107$, $0.0999$ and $0.7907$ respectively, and the corresponding curve is also illustrated in Fig. \ref{fig:recgAccuracy}. 

\section{Conclusion} \label{conclusion}

In this letter, an integrated passive sensing and communication system in 60 GHz band is elaborated, where phased arrays are deployed at both the transmitter and receiver for beamforming to the reference channel and surveillance channel. To demonstrate the performance of this system in motion detection, three types of gestures are made in the surveillance channel and  a dataset of received signals is collected via the above system. Then the ResNet for gesture classification is trained by the dataset. It is shown by experiments that passive sensing in 60 GHz has a good resolution on the micro-Doppler effect of hand gestures as the classification accuracy is greater than 90\%. It is also robust to link blockage as good classification accuracy can be achieved even the NLoS path is used as the reference channel. Finally, an empirical model of classification accuracy is derived from the experiment results via curve fitting.

\bibliographystyle{IEEEtran}
\bibliography{Passive_mmWave_Motion_Detection}

\end{document}